\documentclass[letterpaper, 10pt, conference]{ieeeconf}
\usepackage{mathptmx}
\usepackage[scaled=0.92]{helvet}
\usepackage{courier}
\usepackage{amsmath}
\usepackage{graphicx}
\usepackage{tikz}
\usetikzlibrary{external}

\usepackage{float}
\usepackage{outlines}
\usepackage{caption}
\usepackage[backend=biber,
  style=ieee,
  minnames=1,
  maxcitenames=2, maxbibnames=6,
  doi=false,
  isbn=false,
  url=false,
  eprint=false
]{biblatex}
\DeclareSourcemap{
  \maps[datatype=bibtex]{
    \map{
      \step[fieldset=language, null]
      \step[fieldset=abstract, null]
      \step[fieldset=keywords, null]
      \step[fieldset=file, null]
      \step[fieldset=urldate, null]
      \step[fieldset=note, null]
    }
  }
}
\usepackage{subcaption}
\usepackage{physics}
\usepackage[dvipsnames]{xcolor}
\usepackage{circuitikz}
\IEEEoverridecommandlockouts  
\title{Describing Functions and Phase Response Curves of Excitable Systems}
\author{Robin~Wroblowski$^1$, Rodolphe~Sepulchre$^{1,2}$
\thanks{The research leading to these results has received funding from the European Research Council under the Advanced ERC Grant Agreement SpikyControl n.101054323.}
\thanks{Email: {\itshape robin.wroblowski@kuleuven.be}, {\itshape rodolphe.sepulchre@kuleuven.be}.}
\thanks{$^1$KU Leuven, Department of Electrical Engineering (ESAT), STADIUS Center for Dynamical Systems, Signal Processing and Data Analytics, KasteelPark Arenberg 10, 3001 Leuven, Belgium.}%
\thanks{$^2$Department of Engineering, University of Cambridge, Trumpington Street, Cambridge CB2 1PZ, United Kingdom.}%
}

\begin{document}
\maketitle

\begin{abstract}
The describing function (DF) and phase response curve (PRC) are classical tools for the analysis of feedback oscillations and rhythmic behaviors, widely used across engineering, biology, and neuroscience.
However, these methods are not directly applicable to excitable systems, as they violate the underlying assumptions of harmonic balance and periodicity.
Therefore, the paper proposes a novel approach to extending these classical methods tailored to excitable systems. 
Our methods rely on an abstraction to a discrete-event map from input events to output events. 
The methodology is illustrated on the excitability model of Hodgkin--Huxley. 
The proposed framework provides a basis for designing and analyzing central pattern generators in networks of excitable neurons, with direct relevance to neuromorphic control.

\end{abstract}

\section{Introduction}

The analysis of rhythmic activity in neuronal networks is a central problem both in neuroscience and in neuromorphic engineering. 
Traditionally, tools such as the describing function (DF) and phase response curve (PRC) have been instrumental for engineers and biologists in characterizing and predicting oscillatory behaviors~\cite{ermentrout_oscillator_1990, winfree_geometry_2001}. 

Excitable systems do not oscillate on their own. 
Instead, they only generate action potentials (spikes) in response to triggering stimuli~\cite{izhikevich_dynamical_2006}, greatly simplifying CPG design compared to using autonomous oscillators.
Yet traditional DF/PRC methods, while insightful for oscillator mechanisms and synchronization, cannot be directly applied due to their underlying assumptions.
The describing function method derives from harmonic balance analysis, assuming quasi-harmonic waveforms. 
In contrast, spiking neurons exhibit event-driven behaviors with switch-like transients and fast-slow dynamics~\cite{marder_central_2001}. 
Likewise, phase response curve methods rely on variational methods assuming limit cycle oscillations, rendering them unsuitable for the excitable regime. 

This paper adapts classical DF/PRC methods for event-driven excitable networks, leveraging their strengths while addressing their limitations.
To this end, we consider a reduced model that preserves only the map from input event timings to output event timings.
The event describing function (eDF) quantifies the phase shift between periodic input/output event trains as a function of input period.
Similarly, the event phase response curve (ePRC) quantifies phase shifts between perturbed and nominal event trains as a function of perturbation timing.

Our analysis demonstrates that those adaptations of classical tools preserve their simplicity and insightfulness while showing strong predictive power in excitable networks. 
The approach draws inspiration from recent work by Huo \textit{et al.}~\cite{huo_winner-takes-all_2025}, who showed how winner-takes-all topologies with rebound excitable neurons enable intuitive CPG design.
Ultimately, the framework delivers actionable design principles for rhythmic neuromorphic networks built from excitable components.
\section{Motivation: background \& limitations}

    This section revisits two classical tools for the analysis of nonlinear oscillators: the describing function (DF) and the phase response curve (PRC).  
    Both methods are highly effective when the oscillations are weakly nonlinear or quasi-harmonic. 
    Instead, they face methodological difficulties and lack predictive power when the oscillations are of the relaxation-type.  

    While extensions exist that relax either the harmonic assumptions or the limit-cycle requirement, none simultaneously address both the non-smooth, event-driven nature of spikes and the lack of intrinsic oscillations in excitable systems.
    Our event-based framework exploits these properties, discrete events and forced oscillations, to overcome both limitations at once.

\subsection{Describing functions}

The describing function (DF) method was developed for feedback systems combining linear time-invariant dynamics with static nonlinearities. 
The DF \(N(A,\omega)\) represents the gain and phase shift \(G(A,\omega)e^{j\varphi(A,\omega)}\) induced by a sinusoidal input of amplitude \(A\) and frequency \(\omega\), where both quantities depend on input amplitude and frequency for dynamic nonlinearities~\cite{ghirardo_state-space_2015}.

As a special case of harmonic balance, the DF approximates periodic solutions using only the fundamental harmonic, treating nonlinearities as complex gains. 
This works reliably only when strong low-pass filtering attenuates higher harmonics~\cite{khalil_nonlinear_2014}.

Despite simplifying assumptions, DFs effectively predict oscillation existence and stability in many systems. 
However, relaxation oscillators and neuronal slow-fast dynamics produce abrupt transitions that violate these assumptions. 
These provide only qualitative results~\cite{wang_analyzing_2017, ghirardo_state-space_2015} and motivate alternatives for non-harmonic, event-driven systems.

\subsection{Phase response curves}

The phase response curve (PRC) quantifies how a small perturbation applied to a nominal oscillation affects the phase of the steady-state behavior. 
It describes the phase sensitivity of an oscillator and supports analysis of synchronization and entrainment~\cite{winfree_geometry_2001, izhikevich_dynamical_2006, smeal_phase-response_2010, guckenheimer_nonlinear_1983}. 

PRCs also provide a direct means to analyze and predict phase locking (also under frequency mismatch).
Zero crossings of the PRC indicate potential locked states, and the slope at those points determines their stability.

In addition, beyond descriptive use, PRCs support control design strategies for oscillator synchronization and phase stabilization~\cite{efimov_controlling_2009, qiao_entrainment_2017}. 

Yet, although analytical phase reduction works for simple systems, high-dimensional or strongly nonlinear oscillators require numerical computation.
Moreover, PRC-based analysis degrades in systems with pronounced timescale separations or discontinuities, where switch-like trajectories violate smoothness assumptions~\cite{sacre_singularly_2016}.

\subsection{Limitations}

Both the DF and PRC are not directly applicable to neurons in the excitable regime because of their assumptions: in the excitable regime, neurons are \textit{by definition} not intrinsically oscillating, and the spiking waveform (event) is strongly localized in time, in contrast to harmonic oscillations that are strongly localized in frequency.

Extensions of phase response theory have attempted to relax these constraints.  
Izhikevich~\cite{izhikevich_phase_2000} reformulated phase response theory for relaxation oscillators, and Sacr\'e and Franci~\cite{sacre_singularly_2016} proposed the singular PRC for the limit of strong timescale separation.
Two notable extensions for excitable or nonoscillatory systems are the functional PRC (fPRC) and the isostable response curve (IRC).  
In the fPRC~\cite{cui_functional_2009, sieling_phase_2012}, inputs are generated adaptively from the last output event to measure phase sensitivity in non-oscillatory neurons.
However, this creates an input-output feedback loop (rather than open-loop input-output analysis) and typically employs fixed pulse shapes rather than synaptically filtered events.

In the IRC~\cite{wilson_extending_2015}, isostables (surfaces of equal return rate to equilibrium) characterize how input perturbations affect the spiking waveform during return to rest, whereas we focus on the effects of perturbations on spike generation.

This motivates a new perspective, which we take by focusing on spike timings rather than continuous trajectories (Sec.~\ref{sec:DEES}).
\section{Discrete-Event modeling of excitable systems}
\label{sec:DEES}

    Excitable systems offer key advantages for neuronal modeling (see Huo et al.), including intuitive design and their fundamentally event-driven nature: discrete voltage spikes trigger synaptic currents that propagate through chains of excitable events rather than continuous oscillations.
    Since Hodgkin and Huxley~\cite{hodgkin_quantitative_1952}, neuronal excitability has been modeled by conductance-based circuits (Figure~\ref{fig:HH:circuit}), while simplified models such as integrate-and-fire, FitzHugh-Nagumo, and rate-based models provide computationally and analytically tractable abstractions~\cite{gerstner_neuronal_2014}.
    
    The membrane voltage \(V\) in conductance-based models follows the RC-circuit equation:
    \begin{equation}
            C \frac{dV}{dt} = \sum_{k} I_k, \quad I_k = g_k(t)(E_k - V),
    \end{equation}

    where \(C\) is the membrane capacitance, \(g_k(t)\) are voltage-dependent conductances, and \(E_k\) the reversal potentials.
    
    \begin{figure}[ht]
    \centering
    \begin{subfigure}[b]{\linewidth}
        \centering
        \begin{tikzpicture}
            \draw (4.25, 19) to[capacitor, /tikz/circuitikz/bipoles/length=0.980cm, l_={$C$}, label distance=-0.03cm] (4.25, 17);
            \draw (5.5, 19) to[american resistor, /tikz/circuitikz/bipoles/length=0.840cm, l_={$g_L$}, label distance=-0.01cm] (5.5, 17.5);
            \draw (7, 19) to[variable american resistor, /tikz/circuitikz/bipoles/length=0.840cm, l_={$g_{Na}$}, label distance=-0.05cm] (7, 17.5);
            \draw (8.5, 19) to[variable american resistor, /tikz/circuitikz/bipoles/length=0.840cm, l_={$g_K$}, label distance=-0.05cm] (8.5, 17.5);
            \draw (4.25, 19) -- (8.5, 19);
            \draw (7, 17.5) to[battery1, /tikz/circuitikz/bipoles/length=0.840cm, l_={$E_{Na}$}, label distance=-0.07cm] (7, 17);
            \draw (8.5, 17.5) to[battery1, /tikz/circuitikz/bipoles/length=0.840cm, l_={$E_K$}, label distance=-0.07cm] (8.5, 17);
            \draw (4.25, 17) -- (8.5, 17);
            \draw (5.5, 17.5) -| (5.5, 17);
            \draw (10, 19) to[variable american resistor, /tikz/circuitikz/bipoles/length=0.840cm, l_={$g_{syn}$}, label distance=-0.05cm] (10, 17.5);
            \draw (10, 17.5) to[battery1, /tikz/circuitikz/bipoles/length=0.840cm, l_={$E_{syn}$}, label distance=-0.07cm] (10, 17);
            \draw (8.5, 19) -- (10, 19);
            \draw (10, 17) -- (8.5, 17);
        \end{tikzpicture}
        \caption{Circuit Diagram of the excitable neuron model}
        \label{fig:HH:circuit}
    \end{subfigure}
    
    \vspace{1em}  

    \begin{subfigure}[b]{\linewidth}
        \centering        
        \tikzset{every picture/.style={line width=0.75pt}} 
        
        \begin{tikzpicture}[x=0.75pt,y=0.75pt,yscale=-1,xscale=1]
        
        \draw   (378,118.17) .. controls (378,101.51) and (391.51,88) .. (408.17,88) .. controls (424.83,88) and (438.33,101.51) .. (438.33,118.17) .. controls (438.33,134.83) and (424.83,148.33) .. (408.17,148.33) .. controls (391.51,148.33) and (378,134.83) .. (378,118.17) -- cycle ;
        \draw   (306,100) -- (356,100) -- (356,136) -- (306,136) -- cycle ;
        \draw    (355.83,118) -- (375,117.97) ;
        \draw [shift={(377,117.96)}, rotate = 179.9] [color={rgb, 255:red, 0; green, 0; blue, 0 }  ][line width=0.75]    (10.93,-3.29) .. controls (6.95,-1.4) and (3.31,-0.3) .. (0,0) .. controls (3.31,0.3) and (6.95,1.4) .. (10.93,3.29)   ;
        \draw    (514.6,117.5) -- (514.6,100) ;
        \draw [shift={(514.6,98)}, rotate = 90] [color={rgb, 255:red, 0; green, 0; blue, 0 }  ][line width=0.75]    (7.65,-2.3) .. controls (4.86,-0.97) and (2.31,-0.21) .. (0,0) .. controls (2.31,0.21) and (4.86,0.98) .. (7.65,2.3)   ;
        \draw    (524.43,117.5) -- (524.43,100) ;
        \draw [shift={(524.43,98)}, rotate = 90] [color={rgb, 255:red, 0; green, 0; blue, 0 }  ][line width=0.75]    (7.65,-2.3) .. controls (4.86,-0.97) and (2.31,-0.21) .. (0,0) .. controls (2.31,0.21) and (4.86,0.98) .. (7.65,2.3)   ;
        \draw    (534.01,117.5) -- (534.01,100) ;
        \draw [shift={(534.01,98)}, rotate = 90] [color={rgb, 255:red, 0; green, 0; blue, 0 }  ][line width=0.75]    (7.65,-2.3) .. controls (4.86,-0.97) and (2.31,-0.21) .. (0,0) .. controls (2.31,0.21) and (4.86,0.98) .. (7.65,2.3)   ;
        \draw    (510,117.7) -- (557,117.98) ;
        \draw [shift={(560,118)}, rotate = 180.34] [fill={rgb, 255:red, 0; green, 0; blue, 0 }  ][line width=0.08]  [draw opacity=0] (7.14,-3.43) -- (0,0) -- (7.14,3.43) -- cycle    ;
        \draw    (543.84,117.5) -- (543.84,100) ;
        \draw [shift={(543.84,98)}, rotate = 90] [color={rgb, 255:red, 0; green, 0; blue, 0 }  ][line width=0.75]    (7.65,-2.3) .. controls (4.86,-0.97) and (2.31,-0.21) .. (0,0) .. controls (2.31,0.21) and (4.86,0.98) .. (7.65,2.3)   ;
        
        \draw   (459.5,100) -- (509.5,100) -- (509.5,136) -- (459.5,136) -- cycle ;
        \draw    (262.6,117.5) -- (262.6,100) ;
        \draw [shift={(262.6,98)}, rotate = 90] [color={rgb, 255:red, 0; green, 0; blue, 0 }  ][line width=0.75]    (7.65,-2.3) .. controls (4.86,-0.97) and (2.31,-0.21) .. (0,0) .. controls (2.31,0.21) and (4.86,0.98) .. (7.65,2.3)   ;
        \draw    (272.43,117.5) -- (272.43,100) ;
        \draw [shift={(272.43,98)}, rotate = 90] [color={rgb, 255:red, 0; green, 0; blue, 0 }  ][line width=0.75]    (7.65,-2.3) .. controls (4.86,-0.97) and (2.31,-0.21) .. (0,0) .. controls (2.31,0.21) and (4.86,0.98) .. (7.65,2.3)   ;
        \draw    (282.01,117.5) -- (282.01,100) ;
        \draw [shift={(282.01,98)}, rotate = 90] [color={rgb, 255:red, 0; green, 0; blue, 0 }  ][line width=0.75]    (7.65,-2.3) .. controls (4.86,-0.97) and (2.31,-0.21) .. (0,0) .. controls (2.31,0.21) and (4.86,0.98) .. (7.65,2.3)   ;
        \draw    (256,118) -- (303,118.28) ;
        \draw [shift={(306,118.3)}, rotate = 180.34] [fill={rgb, 255:red, 0; green, 0; blue, 0 }  ][line width=0.08]  [draw opacity=0] (7.14,-3.43) -- (0,0) -- (7.14,3.43) -- cycle    ;
        \draw    (292,117.5) -- (292,100) ;
        \draw [shift={(292,98)}, rotate = 90] [color={rgb, 255:red, 0; green, 0; blue, 0 }  ][line width=0.75]    (7.65,-2.3) .. controls (4.86,-0.97) and (2.31,-0.21) .. (0,0) .. controls (2.31,0.21) and (4.86,0.98) .. (7.65,2.3)   ;
        \draw    (438.83,118.04) -- (458,118) ;
        \draw [shift={(460,118)}, rotate = 179.9] [color={rgb, 255:red, 0; green, 0; blue, 0 }  ][line width=0.75]    (10.93,-3.29) .. controls (6.95,-1.4) and (3.31,-0.3) .. (0,0) .. controls (3.31,0.3) and (6.95,1.4) .. (10.93,3.29)   ;
        
        \draw (331.17,118) node  [font=\footnotesize] [align=center] {Synapse};
        \draw (408.94,118) node   [align=left] {Neuron};
        \draw (448,105) node   [align=left] {V};
        \draw (368,105) node   [align=left] {I};
        \draw (484.5,118) node [font=\footnotesize, align=center] {Event\\ detector};
        \end{tikzpicture}
        \caption{Input-output diagram representation of the discrete-event node. The synapse maps events to currents (E/A), the neuron maps currents to voltages (A/A) and the event detector maps the voltages into events (A/E).}
        \label{fig:HH:tikz}
    \end{subfigure}

    \caption{Excitable neuronal circuit and discrete-event model of fundamental excitable node.}
    \label{fig:HH}
\end{figure}

    Figure~\ref{fig:HH:circuit} shows the electrical circuit model of an excitable neuron joined by a synapse, receiving presynaptic voltage as input and producing membrane voltage as output. 
    Internal currents depend on the neuron voltage, while synaptic currents depend on presynaptic voltage. 
    We choose $E_L = 0$, such that the voltage is expressed relative to the leak reversal potential.   

    For the remainder of the paper, the internal currents correspond to Hodgkin--Huxley ion channels (\(I_L\), \(I_{Na}\), \(I_K\)), while we adopt a generic synaptic conductance model:
    \begin{align}
    g_{syn}(t) &= \bar{g}_{syn} \, h, \\
    \dot{h} &= \frac{\alpha \: (1-h)}{1+\exp\left(\frac{- (V_{pre} - V_{th})}{k}\right)} - \beta h,
    \label{eq:syn_exp}
    \end{align}
    where \(\bar{g}_{syn}\) is the maximal conductance, \(\beta = \tau_{decay}^{-1}\), \(\alpha = \tau_{rise}^{-1} - \beta\), \(V_{th}\) the half-activation constant, and \(k\) the sigmoid slope.

    The discrete-event node shown in Figure~\ref{fig:HH:tikz} defines a continuous-time mapping from presynaptic voltage spike timings to postsynaptic spike timings, and is central to the experiments in this paper.
    Notably, this modeling framework is not limited to Hodgkin--Huxley neurons or specific synapse models. 
    The only requirements are: (1) an event-to-current mapping (E/A) by the synapse; (2) a current-to-voltage mapping (A/A) by the neuron; and (3) a suitable voltage-to-event mapping by the event detector. 
    The choice of event detector must align with the class of events under consideration. 
    For excitable neuron models such as the Hodgkin--Huxley model, a simple threshold detector based on upward crossings of $V_{\mathrm{th}}$ suffices to capture all spiking events.

    To associate a discrete-event model with a continuous-time excitable system, spikes are abstracted as impulses \(\delta(t - t_i)\). 
    This models neuronal communication as a mapping from event sequences to event sequences, consistent with biological spikes triggering synaptic currents.
    It is worthwile to note that the extraction of timing information in this framework is conceptually similar to the projection on a poincar\'e section.
    The proposed representation exclusively considers models in the 1:1 phase-locking mode, in which each input event triggers exactly one output event, creating a one-to-one mapping between event sequences.
    This is a simple, non-restrictive assumption for neurons operating in the excitable regime, as discussed in the next section.

\section{Event describing functions}
    \label{sec:DF}

    The discrete-event modeling framework introduced in Section~\ref{sec:DEES} captures the fundamental event-driven nature of excitable neuronal systems, where spike events, rather than continuous oscillations, govern system dynamics.  
    Within this perspective, we develop a variant of the classical describing function method tailored to event-based input-output signals.  
    This leads to the concept of the \textit{event describing function} (eDF), which characterizes the phase relationship between sequences of input and output events of an excitable system.
    
    Since the input and output signals consist only of unitary events (as discussed in Section~\ref{sec:DEES}), the eDF focuses solely on phase differences. 
    When the input event sequence has a sufficiently large period \(T > T_{\min}\), output events are phase-locked (or \textit{entrained}) to the input events. 
    This allows us to define the eDF as the steady-state event delay \(\Delta(T)\) divided by the input period \(T\), to obtain a relative phase shift:
    \begin{equation}
    \label{eq:eDF}
        \varphi(T) = \frac{\Delta(T)}{T}.
    \end{equation}
    
    This exclusive focus on timing relationships is justified by biologically grounded assumptions that disregard the gain and amplitude dependence of the classical formulation.
    Throughout this paper, signal timing is expressed via the period \(T = \frac{2\pi}{\omega}\), consistent with conventional neuroscience notation.
    
    \subsection{Case study: eDF of a Hodgkin--Huxley neuron}
        \label{sec:DF:case}
        This section analyzes the eDF as a tool to characterize the input-output behavior of the fundamental excitable node within the discrete-event modeling framework.
        The steady-state relative phase shift \(\varphi(T)\) between the event signals is obtained through numerical integration, from which the eDF is constructed by simulating across a range of input signal periods.\\
        
        \begin{figure}[!ht]
    \centering        
        \includegraphics[width=.9\linewidth]{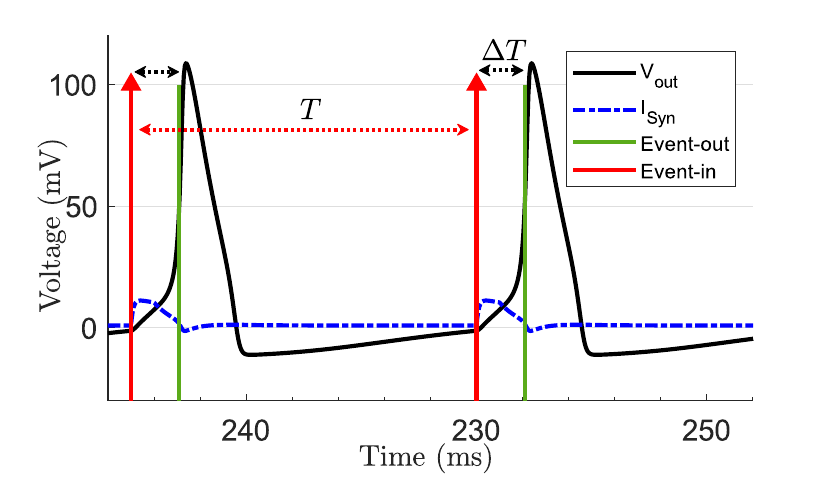}
    
        \vspace{0em}

        \includegraphics[width=.9\linewidth]{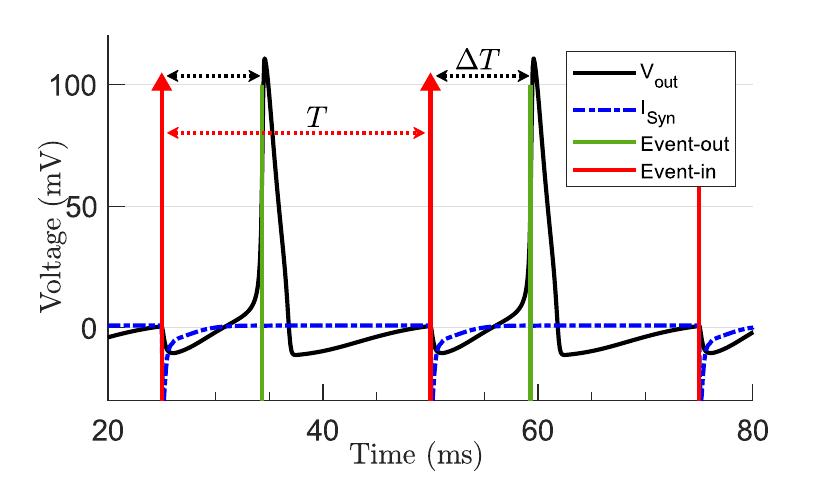}

    \caption{Sample eDF simulation ($\tau_{decay}=1$) showing input events (red vertical arrows), synaptic current (blue), output voltage (black), and output events (green vertical lines). The measured quantity $\Delta(T)$ is computed as the steady-state onset. Top: A small excitatory synaptic current ($g_{syn}=0.3$) elicits a spike soon after. Bottom: A strong and long inhibitory synaptic current ($g_{syn}=5$) elicits a delayed rebound spike. }
    \label{fig:df:examples}
    \vspace{-5pt}
\end{figure}

        \begin{figure}[!ht]
    \centering

        \includegraphics[width=.9\linewidth]{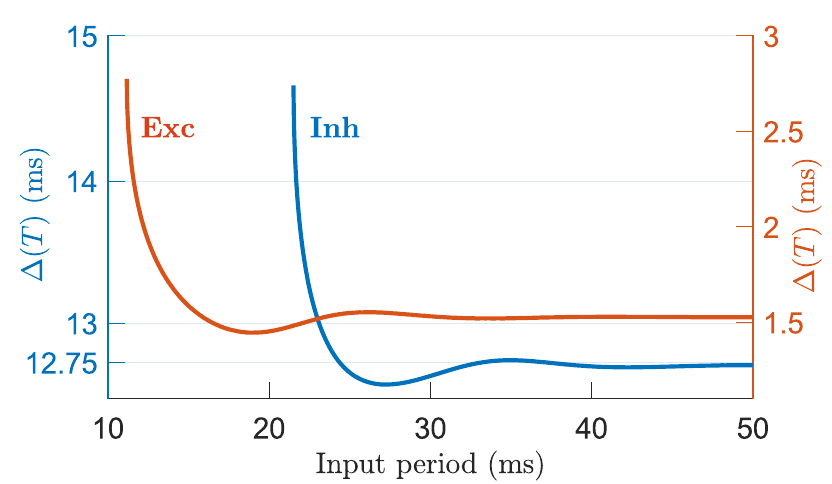}    

        \includegraphics[width=.9\linewidth]{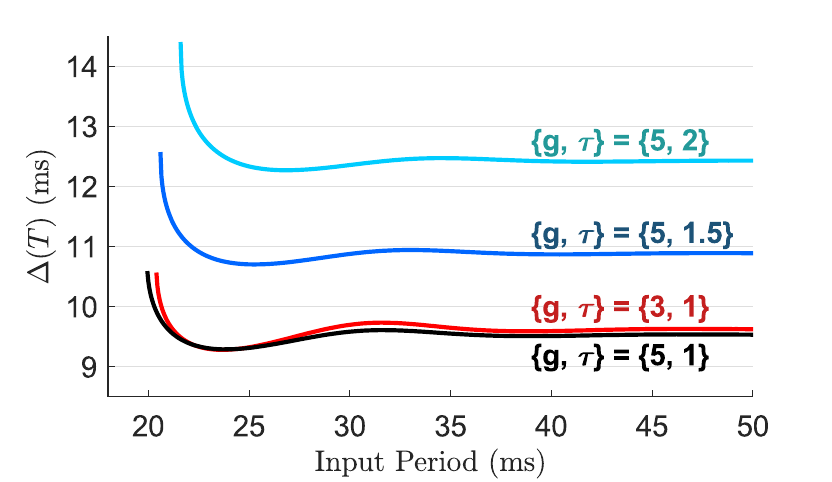}

    \caption{Absolute onset ($\Delta(T)$) curves. Top: comparison of inhibitory (Inh, blue) and excitatory (Exc, orange) node. Bottom: Variation of synaptic parameters for the inhibitory node (varying $g_{syn}$ and $\tau_{decay}$).}
    \label{fig:df:edf}
    \vspace{-15pt}
\end{figure}

        Figure~\ref{fig:df:examples} presents two examples of eDF simulations for an inhibitory node (using a HH neuron) with period \(T=25\) and an excitatory node with \(T=15\) at steady state.  
        The plots display input events (red vertical arrows), synaptic current (blue), output voltage (black), and output events (green vertical lines).  
        Notably, the rebound spike onset is significantly larger, as it occurs only \textit{after} inhibition is released.
        
        Figure~\ref{fig:df:edf} shows the (absolute) onset delay curves ($\Delta (T)$) for the inhibitory and excitatory configuration.
        The upper panel displays the event onset times over a range of input periods.  
        As expected, both remain constant (\(\Delta T(\infty)\)) above a certain `resting period' \(T_r\), corresponding to the total time of activity (event onset plus refractory period).  
        
        The empty region to the left of the minimum period \(T_{\min}\) indicates an absence of 1:1 phase-locking, with distinct implications for the two scenarios: the excitatory node may show other phase-locked modes (e.g., N:M locking) or exhibit \textit{phase slips} where synchrony is interrupted. 
        Inhibitory nodes do not support higher-order locking but do show phase slips and, at high input frequencies, enter sustained inhibition, preventing output events.
        
        Between the minimum input period and the resting period \(T_{\min} < T < T_r\), deviations from the resting onset \(\Delta T(\infty)\) arise due to nonlinear effects like after-hyperpolarization and after-depolarization.  
        These effects vary with neuron model and parameter choices and represent periods when the prior event's influence persists into the next input.
        
        The lower panel of Figure~\ref{fig:df:edf} shows the onset delay curves of an inhibitory node for varying synaptic decay constants \(\tau_{decay}\) and synaptic strengths \(g_{syn}\).  
        Changing \(g_{syn}\) only slightly influences \(T_r\), \(T_{min}\), and \(\Delta T(\infty)\), while moderately affecting the severity of the nonlinear after-spike effects.  
        In contrast, varying \(\tau_{decay}\) causes clear horizontal and vertical shifts in the onset curves.
        Consequently, modulation of the synaptic decay time constant offers an effective way to adjust the location of the eDF curve in excitable nodes.        
   
    \subsection{Predicting ring network oscillations}
        
\begin{figure}[!t]
    \centering    
     
    \includegraphics[width=.9\linewidth]{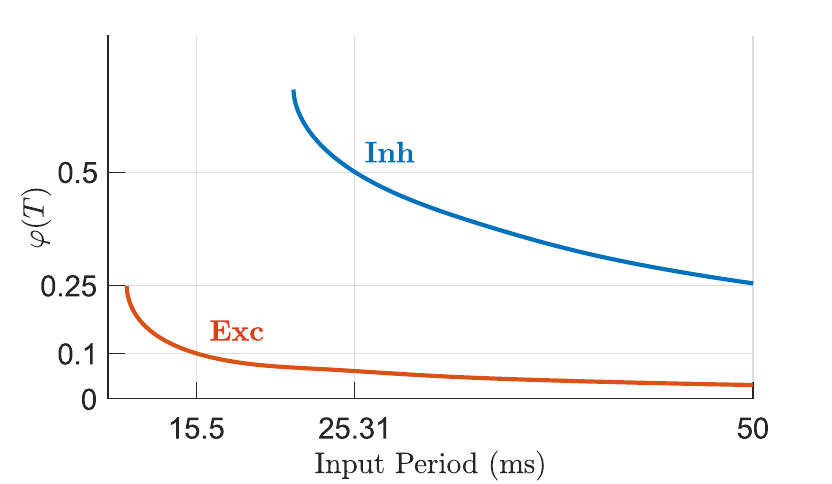}    

    \caption{eDF curve for inhibitory and excitatory node: relative event onset in function of input period enables graphical prediction of network oscillation period and existence.}
    \label{fig:df:edf_network}
\end{figure}

        A straightforward application of the eDF is in predicting network oscillation existence and period: A ring network of $N$ excitable nodes permits a rhythm with period $T$ if the relative phases of its components sum to $1$, i.e.
        \begin{equation}
        \label{eq:ring_stability}
            \sum^N_{i=1} {\varphi_i(T)} = 1,
        \end{equation}
        where the relative phase of excitable system $i$ at input period $T$ is represented by $\varphi_i(T)$.
        In the case of homogeneous nodes in the ring network, the eDFs in Figure~\ref{fig:df:edf_network} enable simple graphical predictions of network oscillation period.
        For instance, we find that the inhibitory eDF curve has an intersection at $\varphi_i(T_N) = 0.5$, which indicates the existence of a stable two-node ring network oscillation at the period of intersection.
        This recurrent connection of two inhibitory nodes corresponds to the half-center oscillator (HCO), which is a fundamental motif in neuroscience and biology, known for its robust rhythmic behavior.
        The excitatory eDF, however, does not intersect the value $0.5$, and thus two-node ring network oscillations cannot exist, although other stable networks may be found for larger rings.\\
        
        The same approach applies to heterogeneous ring networks: summing the individual eDFs and finding the unit intersection (Equation~\ref{eq:ring_stability}) predicts possible oscillation periods.  
        This method enables constructing ring networks that combine inhibitory and excitatory nodes to achieve a desired total period, while parameter variation, as illustrated in Figure~\ref{fig:df:edf}, offers precise control and fine-tuning of the eDF curves to adjust network dynamics.

        The proposed framework can now be validated against neuronal network simulations that solve the underlying continuous-time ODEs.
        The predicted network periods prove to be relatively good estimates, as shown in Table~\ref{tab:DF}.
        The remaining errors can be attributed to the mismatch between the abstraction of the eDF's input impulses and the continuous spiking waveforms of the ODEs.

\begin{table}[!t]
    \centering
    \begin{tabular}{lll}
        Ring type                      & Prediction eDF          & Simulation                \\ \hline
        \multicolumn{1}{|l|}{I-I} & \multicolumn{1}{l|}{25.31} & \multicolumn{1}{l|}{25.61} \\ \hline
        \multicolumn{1}{|l|}{I-I-I-I} & \multicolumn{1}{l|}{50.93} & \multicolumn{1}{l|}{51.32} \\ \hline
        \multicolumn{1}{|l|}{E- x10} & \multicolumn{1}{l|}{15.50} & \multicolumn{1}{l|}{15.49} \\ \hline
    \end{tabular}
    \caption{Predicted and simulated network period (ms) for various configurations. Inhibitory nodes have $g_{syn}=5$ and excitatory nodes have $g_{syn}=.3$.}
    \label{tab:DF}
    \vspace{-20pt}
\end{table}

        If the eDF is monotonically decreasing, it implies that there can be at most one intersection and thus at most a single unique solution for the network oscillation period.
        This uniqueness is closely related to the stability of the solution: a monotonically decreasing eDF curve functions as an indicator for local stability.
        Such relations between monotonicity and stability have been central in the work of Angeli and Sontag~\cite{angeli_detection_2004, angeli_monotone_2003}.
        While this property holds for the Hodgkin--Huxley-type systems considered here, it may not generalize to all excitable or neuronal models (especially those with more prominent after-spike nonlinearities).
        Note that, since the event delay $\Delta(T)$ of an excitable system converges to a fixed value \(\Delta T(\infty)\), the eDF necessarily vanishes as $\lim_{t\to\infty} \varphi(t) = 0$.

\section{Event phase response curves}
    \label{sec:PRC}
    In the previous section, we developed the event describing function (eDF) as an input-output framework to characterize the frequency-dependent response of excitable systems.  
    Here, we take a complementary perspective and focus on perturbation sensitivity.
    Specifically, we extend the classical phase response curve (PRC) concept to event-based systems, introducing the event phase response curve (ePRC).  
    This framework captures how periodic synaptic perturbations affect the timing of the periodically forced discrete-event system, providing a relevant insights into its interaction behavior.\\

    \subsection{Event phase response curves (ePRC)}
    \label{sec:PRC:ePRC}
    
        We propose the event phase response curve (ePRC) as an event-based analogue of the classical PRC for excitable systems.
        The ePRC quantifies the time shift induced by a periodic synaptic perturbation on a nominal periodic event oscillation (of the same period). 
        Periodic perturbations are applied through a synapse, rather than directly perturbing the trajectory, to account for effects from the synaptic parameters and providing a biologically realistic description of network interaction similar to the spike time response curve (STRC).

        The classical PRC measures the phase shift between a nominal oscillation and its perturbed (steady-state) counterpart.
        Analogously, the ePRC of an excitable system is defined with respect to a nominal event-based oscillation caused by a nominal periodic input event sequence.
        The nominal event delay of the system is characterized by its describing function (see Section~\ref{sec:DF}). 
        Perturbations are introduced at delay \( t_p \) within each period of the nominal oscillation, thereby (possibly) affecting the timing of the output event. 
        Consequently, the ePRC defines the relationship between the perturbation timing and the resulting output time shift.
        
        In contrast to classical PRC analysis, which assumes weak coupling, our framework has broader applicability. 
        Strong perturbations generally induce large phase shifts (resets) due to neuronal refractoriness and the excitable regime. 
        Early perturbations preceding the nominal input, however, may trigger additional output events and disrupt 1:1 locking.

        Since the nominal oscillation is purely exogenous, a single perturbation's time-shifting effect does not propagate indefinitely.
        The subsequent nominal input event remains unaffected, necessitating periodic perturbations instead.
        An alternative approach would be to define the nominal oscillation as a recurrent loop where input timings are defined as a delay after the previous output spike (thus allowing single perturbation effects to persist).
        However, this adaptive approach sacrifices the input-output simplicity of fixed input timings, so we retain the original definition with periodic perturbations.

        The ePRC framework enables the prediction of interaction behavior of event-based network oscillations of period $T_N$.
        When $T_N$ is sufficiently large, each input event is isolated, allowing the neuron to return to rest before the next event (as discussed in Section~\ref{sec:DF:case}).
        In this case, the ePRC may be computed from a single nominal event (triggered by a single input event), greatly reducing computational load.

    \subsection{Case study: ePRC of a Hodgkin--Huxley neuron}
    
\begin{figure}[!t]
    \centering        
    \includegraphics[width=.87\linewidth]{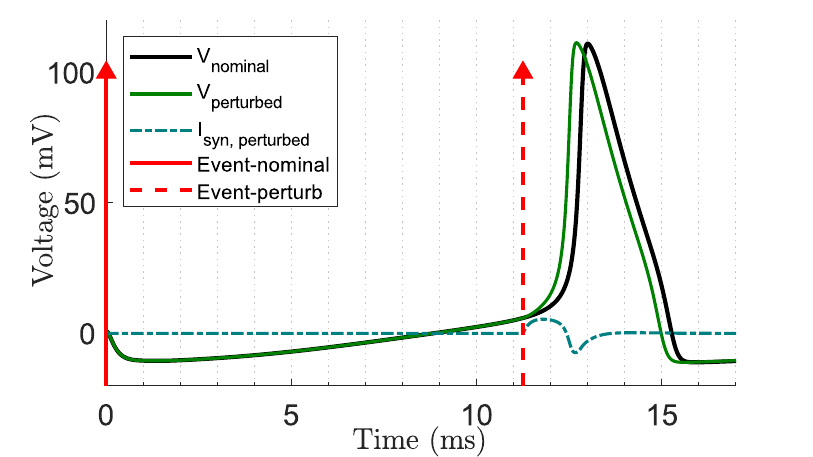}
    \caption{Phase advance caused by an excitatory perturbation on a nominal inhibitory rebound spike. The nominal input event (solid red) elicits a nominal rebound spike (black). The perturbation input (dashed red) causes a change in synaptic current (dashed teal), resulting in an advanced output spike (green).}
    \label{fig:prc:rebound:example}
    \vspace{-10pt}
\end{figure}

\begin{figure}[!t]
    \centering        

    \includegraphics[width=.9\linewidth]{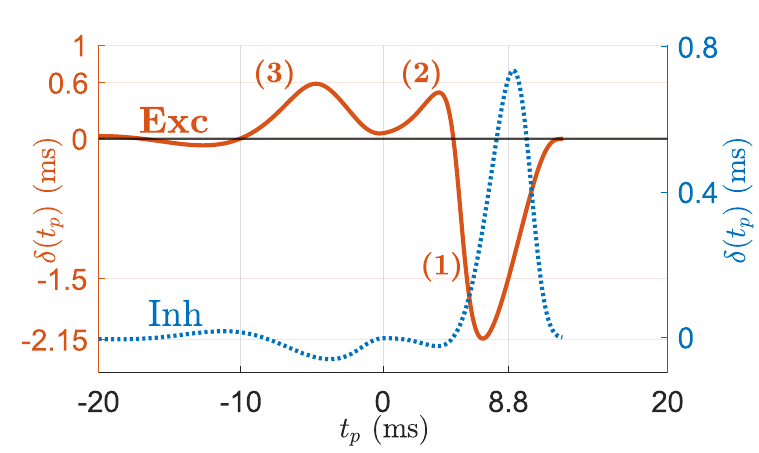}

    \caption{ePRCs for excitatory (orange) and inhibitory (dotted blue) perturbations on a nominal inhibitory rebound spike (on separate axes).}
    \label{fig:prc:rebound:curves}
    \vspace{-15pt}

\end{figure}

        We next apply the ePRC method to the inhibitory discrete-event node as introduced in Section~\ref{sec:DEES}, using a HH neuron, with perturbations delivered via excitatory or inhibitory synapses.
        Figure~\ref{fig:prc:rebound:example} presents a sample instance of the ePRC simulation for the excitatory perturbation, where the perturbation induces an shift in onset delay (\(\delta(t_p)\)).
        Repeating this simulation over a range of \(t_p\) yields the ePRCs in the top panel of Figure~\ref{fig:prc:rebound:curves}. 
        Throughout this study, we adopt the standard sign convention where a positive PRC value denotes a phase delay, and define the perturbation time \(t_p = 0\) when the perturbation coincides with the nominal input event.
        The variable $t_p$ is not restricted to positive values; negative $t_p$ correspond to perturbations delivered before the nominal input event, effectively preconditioning the internal states of the neuron.\\

        The excitatory ePRC exhibits three zero \textit{crossings} (we neglect small values for \(t_p < -15\) and include the rightmost zero), indicating potential phase-locking equilibria. 
        The middle equilibrium (\(t_p \approx 5\)) is unstable due to its negative slope, whereas the far-right equilibrium coincides with the baseline output time and thus later perturbations do not affect the timing at all.
        
        The ePRC regions reflect biophysical effects: (1) delayed excitation aids inhibitory recovery, advancing the spike; (2) excitation near inhibition onset reduces Na-current deinactivation, delaying it; and (3) early excitation activates K-current, further delaying recovery.\\   
        
        By vertically shifting the ePRC to mimic frequency mismatch between the forcing input and the intrinsic network period, we can explore entrainment properties~\cite{forger_biological_2017}. 
        A downward shift (lower forcing frequency) introduces a new stable equilibrium near \(t_p \approx 0\), removes the in-phase equilibrium, and enlarges the basin of attraction for the leftmost equilibrium. 
        Conversely, an upward shift (higher forcing frequency) leaves the rightmost equilibrium as the only stable point. 
        This indicates that inhibitory network oscillations entrain in-phase and robustly when the forcing frequency exceeds the intrinsic network frequency. 
        The effective frequency mismatch is limited by the peak amplitude (near \(t_p \approx 7\)), since excessive mismatch, and thereby a large vertical shift, can make all zero crossings vanish.\\
        
\begin{figure}[!h]
    \centering

        \includegraphics[width=.9\linewidth]{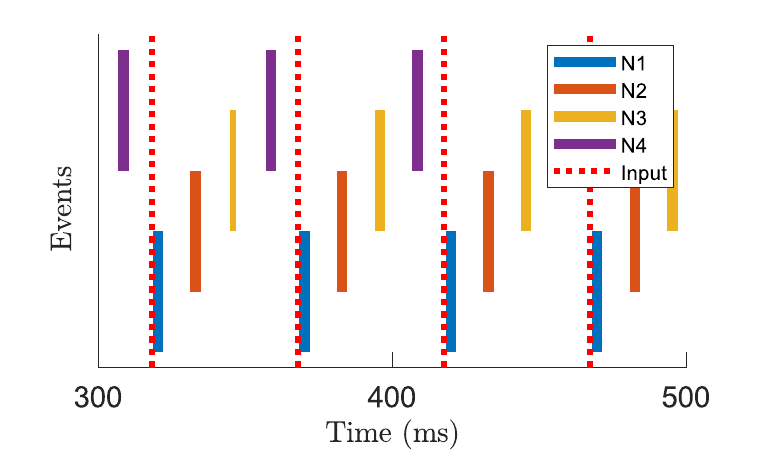}

    \caption{Experimental validation of the ePRC in a ring network of four inhibitory nodes (with network period $T_N=51.2$). Periodic excitatory perturbations are applied to neuron N1 at a lower period ($\Delta (T) = -1.5$). As expected, the network is entrained to the perturbations.}
    \label{fig:prc:entrainment}
    \vspace{-10pt}
\end{figure}

        In Figure~\ref{fig:prc:entrainment}, a ring of four inhibitory nodes with an intrinsic period \(T_{N}=51.2\) is driven by excitatory inputs at both higher and lower frequencies. 
        As expected, the network remains phase-locked to the forcing: the ePRC predicts a lag of \(t_p \approx 8.8\) for a period mismatch of \(\Delta T = -1.5\), while the simulation yields a lag of \(t_p = 8.1\). 
        The same analysis holds for other types of perturbations, nominal inputs and period mismatches.

        Overall, these results indicate that rebound spiking neurons entrain robustly to inhibitory perturbations at lower or matching frequencies (out of phase) and to excitatory perturbations at higher frequencies (in phase).
        Although not shown in this paper, excitatory neurons have similar entrainment properties.
        This novel approach could offer practical guidance for the design of the interaction properties of phase-based controllers.

\section{conclusion}
This work has introduced a novel framework extending classical describing function and phase response curve methodologies to excitable neuronal systems. 
By focusing on event-based input-output characteristics of synapse-neuron nodes, the proposed event describing function (eDF) and event phase response curve (ePRC) provide valuable tools to analyze and predict rhythmic dynamics in excitable networks. 

Extending the analysis to burst-excitable systems and more complex CPG architectures remains an important direction for future research.
We envision that these developments will facilitate the design and control of biologically inspired rhythmic circuits for robotics and neuromorphic engineering.

\begin{tiny}
    \printbibliography
\end{tiny}

\end{document}